# INTERNAL FRICTION MEASURES TO STUDY PRECIPITATES FORMATION IN EP AND N-DOPED BULK Nb FOR SRF APPLICATIONS

Tiziana Spina, Alexander Romanenko, Anna Grassellino

## ABSTRACT

Main focus of this study is the investigation of thermodynamics phenomena responsible for the High Field Q Slope (HFQS) in SRF cavities by Internal Friction (IF) measurement. Mechanical spectroscopy is, indeed, a well-established technique to study precipitate formations in BCC materials and several works on the effects of impurities as N and O on the Snoek peak have been published so far and will be taken as reference to explain the mechanisms behind the observed dissipation effects. Internal Friction measurements were performed in Belgium at IMCE on Nb rectangular shape samples with different RRR values prepared at Fermilab by using Electro Polishing (EP), N-doping and heat treatments in order to reproduce the same conditions during the standard treatments applied on bulk Nb SRF cavities.

From IF spectra, the H trapping mechanism by interstitial atoms (N and O and/or vacancies, depending on the purity level, RRR) can be easily recognized leading to results that perfectly corroborate previous findings on Q-disease, HFQS and RRR phenomena.

## INTRODUCTION

Nowadays the debate on the main players responsible for the High Field Q Slope (HFQS) in Niobium bulk RF cavity is still open and constitute one of the major topics for the scientific community. Indeed, higher performance in SRF cavities at high fields is advantageous for reducing accelerator length and consequently for cryogenics cost saving.

For a severe Q-disease, Q values fall to $10^7$ above a few MV $m^{-1}$. Below 10 at.% concentration, H will remain harmlessly in the disordered α-phase at 300 K. Nb–H phases harmful to superconductivity form between 100 and 150 K, when the H concentration is greater than 100–200 at. ppm. The sharp Q drop for a cavity indicates that Nb–H regions are initially superconducting at low fields (<5 MV m−1) and become normal at higher fields [Padamsee 2008].

Large non-superconducting hydrides were identified as an origin of Q-disease in SRF Nb bulk cavities. By confocal optical microscopy Barkov et al. [2014] observed the formation of micro-hydrides in H-loaded bulk Nb during cool down and their dissolution during warm-up

As reported in [Bonin, 1991], low RRR Nb does not show Q-disease because interstitial impurities (such as oxygen) in Nb serve as trapping centers for hydrogen, preventing hydrogen mobility and hydride growth. Similarly, vacancies and dislocations are also effective traps for hydrogen as also reported by Cizek [Cizek, 2009] that observed by PAS, XRD and TEM irradiation-induced vacancies surrounded by hydrogen. A recent electron irradiation study on H-loaded Nb bulk samples at Fermilab [Spina 2021] by optical confocal microscopy proves that hydrides growth can indeed be inhibited by such treatment.

It has been found that the best cure for the Q-disease is H degassing at 800°C for 3 hours under high vacuum pressure (>$10^{-6}$ Torr), or at 600 °C for 10 hours, if the decrease in yield strength from 800 °C heat treatment cannot be tolerated for other practical reasons [Padamsee 2008].

T-map measures show that cavities characterized by HFQS behavior have a non-uniform distribution of heating along the equator suggesting that some regions (hot) have higher losses than others. In particular, the first experimental evidence of hydrides formation in the two hottest spots of cavity and their correlation with HFQS was found by Romanenko et al. [2012].

An extensive study performed on cut-outs from SRF cavities revealed also the formation of near-surface low temperature nanoscale niobium hydrides phases in cavities with HFQS behavior [Trenikhina, 2015]; recently, nano-hydrides formation was also observed by cryo AFM measurements at Fermilab [Sung 2019].

A theory based on the presence of proximity-coupled niobium hydrides has been proposed by Romanenko et al. [2013] to explain the mechanism for the occurrence of HFQS. They claim that HFQS can be attributed to the breakdown of the proximity effect acting on niobium hydrides in the near surface layer; in addition, the authors suggest that this model can be also applied for a quantitatively description of Q-degradation due to large niobium hydrides (Q-disease).

The cure for the HFQS has been found empirically by baking the cavity at 100-120°C/28-48h (mild bake).

In particular, HFQS is completely removed for EP cavity while only an improvement is observed for BCP cavity. A possible explanation for this different behavior for EP and BCP cavities under mild baking can be assigned to the vacancy enhancement induced by the EP process; vacancies may trap hydrogen thus preventing the deleterious effect of hydrides formation [Romanenko 2013A].

H. Safa in 2001 [Safa 2001] has suggested that an "oxygen pollution layer" underneath the oxide gets diluted/diffused away by the 120°C baking and could be responsible of the HFQS removal during mild baking.

Recently, 120°C baked cut-out TOF-SIMS studies at Fermilab [Romanenko 2020] revealed O diffusion from "an infinite source" confirming the idea of O as the main player in the 120°C bake treatment instead than vacancies as previously reported [Romanenko 2013A]. This model is in perfect agreement with the hydride model [Romanenko 2013B] for the HFQS emergence indeed it is thought that the injection of O near surface has the double effect to trap hydrogen thus inhibiting hydride formation and reduce the mean free path.

After the N-doping discovery, it was also observed that interstitial N can effectively mitigates HFQS (but not remove) especially when the cavity is submitted to the thermal process with nitrogen at 120°C/48 h after an initial degas at 800°C/3 h (N-infusion) [Grassellino 2015]. A recent review on N-doping and infusion discovery can be found in [Pushpati, 2020].

In 2018 Garg et al. [Garg 2018], by first principles calculations, confirmed that the presence of nitrogen during processing plays a critical role in controlling hydride precipitation and subsequent SRF properties.

In the same year, Grassellino and coworkers [Grassellino 2018] discovered a new surface treatment based on an intermediate low T baking applied just before the standard mild bake allowing unprecedented accelerating fields till 49 MV/m in TESLA-shaped cavities, in continuous wave (CW)). For this treatment the niobium surface has been submitted to electropolishing (EP) followed by a modified mild bake, i.e. adding to the standard mild bake a step at 75°C for few hours, before the 120°C vacuum bake. The authors suggest that a possible reason behind the observed improvement in cavity performance at high accelerating field can be the suppression of nano-hydrides via defect trapping, with defects-H complexes forming at the two different temperatures.

It follows that in order study the mechanisms behind the HFQS phenomena, the effects of interstitials atoms as O, N or/and vacancies on the precipitate's formation need to be explored and represent the main subject of the present work.

Mechanical spectroscopy, based on the measure of Internal Friction (IF) peaks, is a very sensitive technique to detect the formation of hydrides in BCC metals as Nb (the precipitation peak [Yoshinari 1982]) and to understand the interaction with interstitials atoms (Snoek peak) [Weller 2006]. Thus, to study the precipitate formation in SRF bulk Nb cavities during thermal cycles, IF measurements have been performed on EP and N-doped bulk Nb samples to mimic the behavior of cavities characterized by HFQS.

## A NEW TECHNIQUE TO DETECT HYDRIDES FORMATION IN EP AND N-DOPED Nb BULK SAMPLES

Internal friction is the measure of the dissipation of mechanical energy inside a solid material once submitted to a time-dependent load within the elastic deformation range [Nowick and Berry 1972].

Different materials defects such as dislocations or vacancies as well as interstitial atoms can contribute to an increase in the internal friction between the vibrating defects and the neighboring regions. Three are the main mechanisms that can produce anelastic damping in a solid: the diffusive motion under stress of point defects (Gorsky relaxation), the motion of dislocations (Snoek relaxation), and the motion of grain boundaries or other interfaces (relaxation due to H including the precipitation or α peak). An extensive study of all possible relaxation mechanisms can be found in [Blanter 2007] while here the main attention will be paid on the analysis of precipitation (α) and Snoek peaks due to N and O contaminations respectively.

Owing to the large variety of phenomena, materials, and related microscopic models, a correct interpretation of measured internal friction spectra is often difficult. An efficient use of mechanical spectroscopy may then require both a systematic treatment of the different mechanisms of internal friction and anelastic relaxation and a comprehensive compilation of experimental data in order to facilitate the assignment of mechanisms to the observed phenomena.

In this study, experimental data are compared with literature results to explore hydrides formation in EP and N-doped Nb bulk samples thus to identify the main players involved in the HFQS behavior in SRF accelerating cavities.

### Experimental set-up: Impulse Excitation Technique

Internal Friction measures for this work were performed in Belgium at IMCE. This facility was chosen because capable to induce excitation frequencies of the order of kHz in the appropriate temperature range at which the precipitation peak (α-peak) is expected to appear below room T (see for example [Yoshinari and Koiwa, 1988]). It is based on the Impulse Excitation Technique (IET) [Roebben 1997], i.e. a non-destructive material characterization technique to measure the resonant frequencies from which the internal friction of predefined shapes samples (rectangular bars, cylindrical rods and disc shaped samples) can be calculated.

The measurement principle is based on tapping the sample with a small projectile and recording the induced vibration signal with a microphone. Then, the acquired vibration signal in time domain is converted by a fast Fourier transformation to frequency domain. The material damping or internal friction is obtained by the decay of the vibration amplitude in the sample in free vibration as the logarithmic decrement [Roebben 1997].

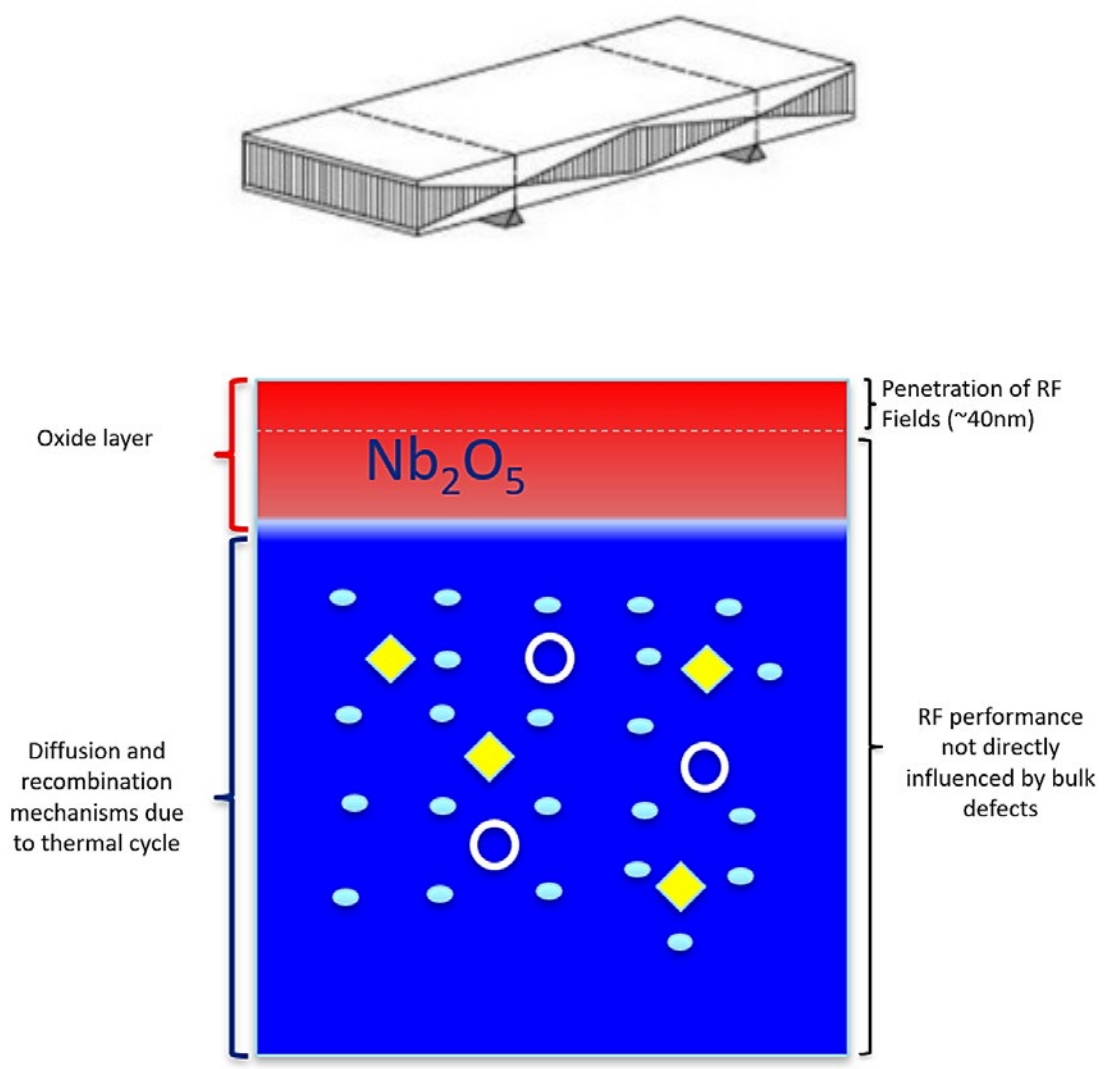

**Figure 1** Bulk vs layer: (top) vibration induced into the bar during IF measurement (bulk); (bottom) regions of interest in Nb samples for SRF applications.

As shown in Figure 1 (top), the entire sample volume contributes to the IF spectra while RF fields occupy only the first ~40 nm (over 3mm of total thickness) of the inner cavity surface; thus, can raise the question if such IF measurements can really be representative for the description of the effects observed in SRF cavities where only the first 100um are important. Indeed, we cannot forget that surface mobility is closely related to diffusion phenomena in the bulk of solids, which have been extensively investigated in [Manning 1968]. Thus, the surface-to-bulk diffusion of interstitial atoms in metals is important in order to study surface phenomena under thermal cycling and thus responsible for the HFQS in SRF cavities.

**Table 1** Nb samples for Internal Friction studies: 2 EP and 2 N-doped

| Name | Treatment | RRR |
|---|---|---|
| EP High RRR [H4] | 120µm EP(20C)+HT(800C/3h) | 300 |
| EP Low RRR [L4] | 120µm EP(20C)+HT(800C/3h) | 20 |
| N-dop High RRR[H2] | 120µm EP(20C)+HT(800C/3h)+ 2/6 N-doping +10µm EP(20C) | 300 |
| N-dop Low RRR[L2] | 120µm EP(20C)+HT(800C/3h)+ 2/6 N-doping +10µm EP(20C) | 20 |

Four rectangular bulk Nb samples with fine grains are described in Table 1. In particular, *High RRR* and *Low RRR* samples are submitted to standard treatment used in Nb bulk RF cavity:

- bulk EP treatment: 120µm EP at 20°C followed by heat treatment at 800°C for 3 hours
- N-doped 2/6 treatment: after the same treatment applied on EP samples (i.e. 120µm EP at 20°C followed by heat treatment at 800°C for 3 hours) N is injected at 25mTorr for 2 minutes and let it diffuse for 6 minutes and finally a light EP is used to remove 10µm from surface (i.e. oxide layer formed during the N doping treatment).

These samples are representative of cavities characterized by HFQS (EP cavity) and anti-Q slope (N-doped cavity) thus can be used to investigate on the main players responsible for such behaviors.

IF Measurements have been performed in flexural mode at a frequency of 2.9kHz and temperature was changed between -80°C and 800°C with a rate of 0.3°C/min in order to cover the range of interest to study the precipitates formation in bulk Nb samples.

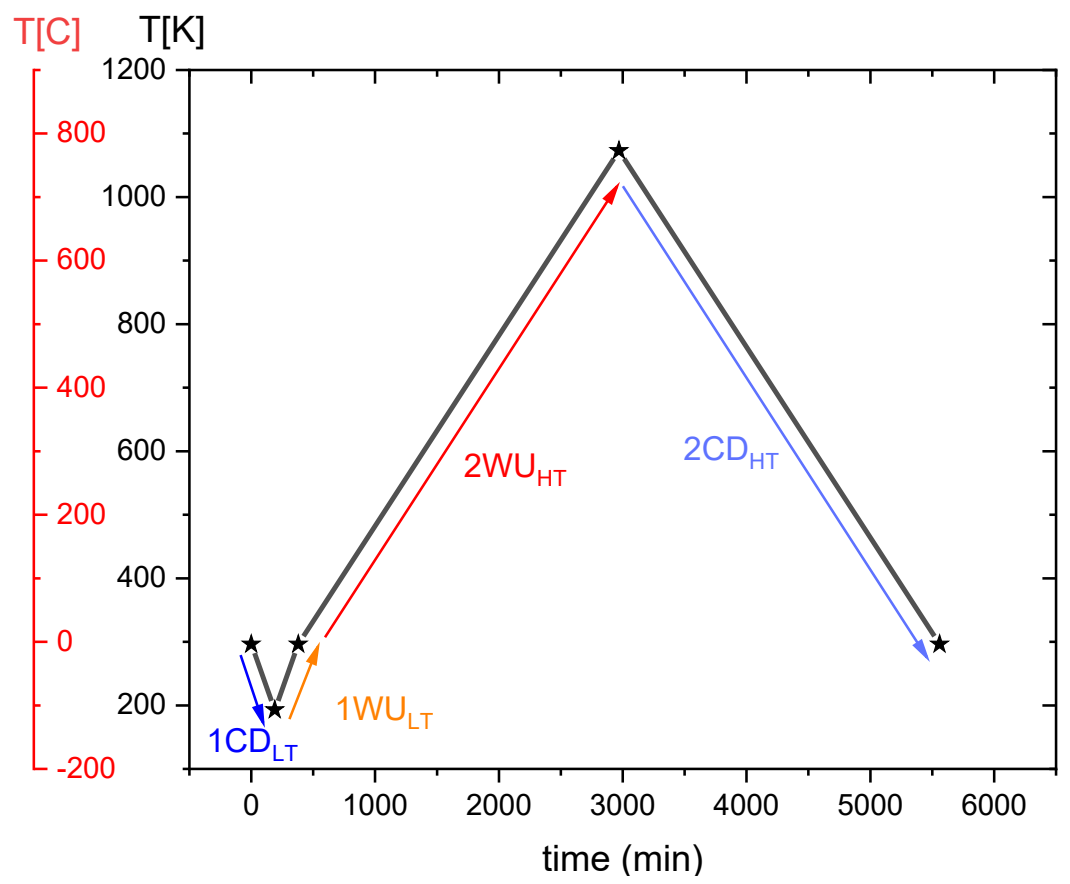

**Figure 2** Thermal cycle

As shown in Figure 2, each thermal cycle consists of 2 warm-ups (WU) and 2 cool-downs (CD): i.e. first CD at low T (from RT to -80ºC) ($1CD_{LT}$) followed by WU at low T (from -80ºC to RT) ($1WU_{LT}$) then the WU continues to reach higher T ($2WU_{HT}$) going from RT to 800ºC and finally CD to go back to RT ($2CD_{HT}$).

Due to the lack of High Vacuum furnace typically used during SRF cavity treatments, an Argon flow has been turned on during High Temperature (HT) warm-up and also a Ti powder source has been introduced to prevent oxidation.

## RESULTS AND DISCUSSION

*EP samples: High vs. Low RRR*

Oxygen, nitrogen, hydrogen, and carbon are the main interstitial dissolved impurities acting as scattering centers for unpaired electrons and reducing the RRR in SRF cavities. As reported by Singer [Singer 2011], the influence of hydrogen on the RRR is not so significant, but the hydrogen content should be kept low (less than 2 µg/g) in order to prevent hydride precipitation in high-RRR cavities under certain cool-down conditions (hydrogen Q disease) [Padamsee 2008]

Thus, EP samples with different RRR values are submitted to thermal cycles (Figure 2) in order to study the thermodynamics of hydrides formation as a function of impurity concentration (inversely proportional to the RRR value).

As shown in Figure 3, multiple internal friction peaks appear during cooling and heating cycles suggesting a thermally activated relaxation process. The two temperature ranges of interest for SRF cavity, i.e. low T and mild T, are investigated. It is found that in the low T regime, three peaks appear between 190 ÷ 300K while at mild T (about 520K) an additional peak raises and, as described in the next section where data at higher T are available, it is due to the Snoek relaxation induced by interstitial oxygen atoms (Nb-O).

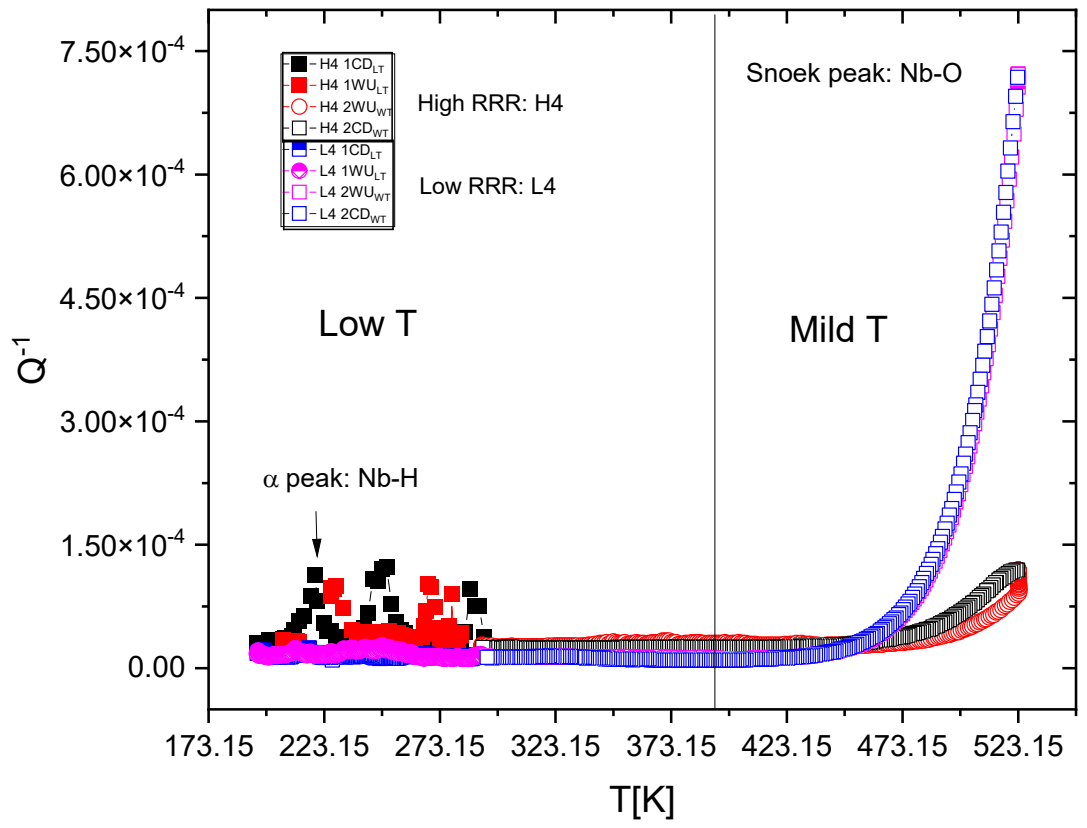

**Figure 3** Internal friction peaks at low and mild T to investigate precipitates and oxide formation in EP samples with high and low RRR

At low T, the three peaks can be assigned to internal mechanisms responsible for hydrides formation/dissolution processes (Nb-H). In particular:

- at **220K** the peak can be regarded as the precipitation peak (or α peak, Nb-H) also known as the Hydrogen Cold Work Peak

(HCWP) in [Mazzolai, 1969; Verdini 1971] and attributed to relaxation process involving dislocation formed around precipitates in analogy on what found at high frequencies (~kHz) by [Yoshinari 1980];

- at **250K** and **290K** the two peaks can be identified as β peaks similarly to those observed in cold-worked niobium samples by Stanley and Szkopiak [Stanley 1967] and possibly caused by relaxation mechanisms involving point-defect complexes.

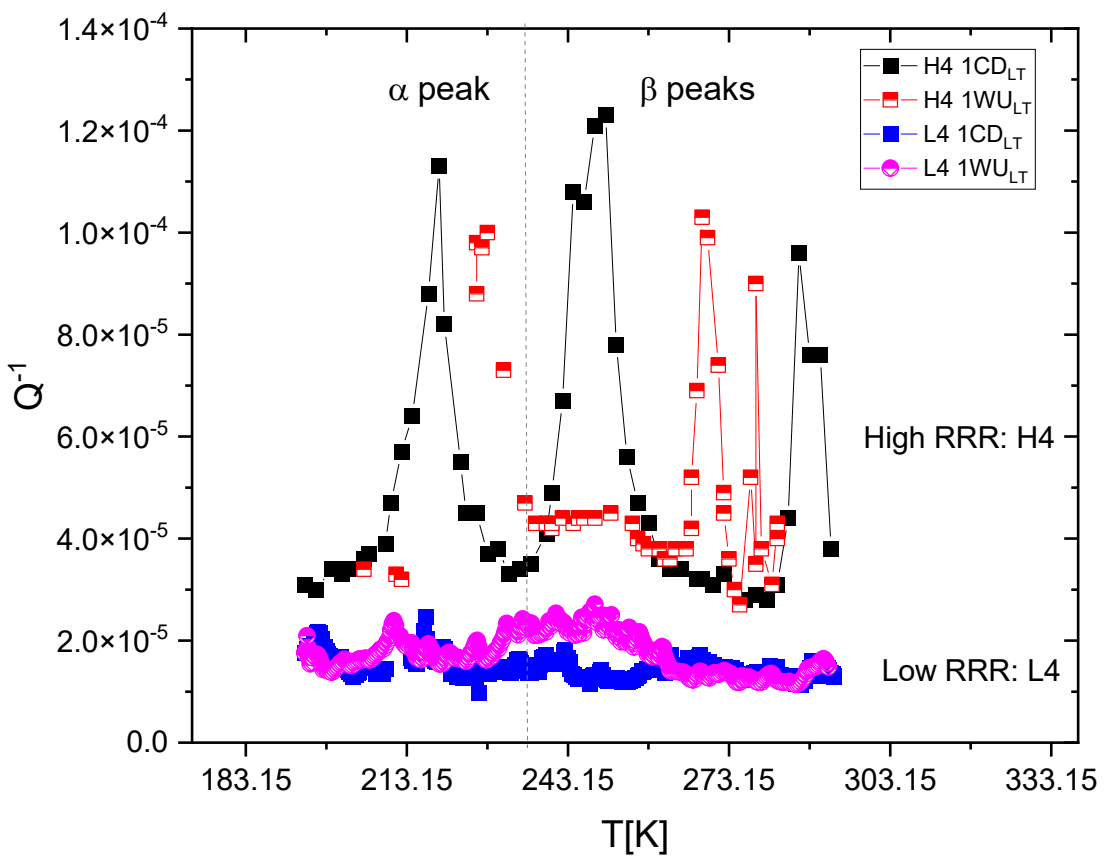

**Figure 4** Comparison High RRR and Low RRR EP Nb samples – precipitation peak disappears in low RRR samples

As shown in Figure 4, at low temperature, High RRR samples are characterized by larger damping than Low RRR and such precipitation peak intensity reduction is assigned to hydrides suppression. Indeed, in low RRR Nb sample, the higher number of trapping centers (either vacancies and/or interstitials atoms as O or C) is responsible for preventing hydrides formation. This result is in perfect agreement with Natsik et al. [Natsik, 2001] that observed a shift in the peak location at higher temperatures and also an increase in its width and amplitude by increasing the purity of the samples (i.e. by increasing RRR values).

In addition, large hysteretic behavior of the alpha peak in High RRR sample during CD and WU can be ascribed to plastic strain introduced into the lattice by hydrides precipitation.

This behavior was also studied by Yoshinari and Koiwa [Yoshinari 1982 I] that measured on thermal cycling a rapid decrease in the height of the precipitation peak and an increase of the precipitation temperature.

Unfortunately, no direct correlation between RRR values and EP SRF cavity performance has been found yet, on the other hand HFQS can be regarded as a Q-disease miniaturized due to the formation of smaller hydrides (nano-hydrides), [Romanenko, 2012]; then, these results are in perfect agreement with the disappearance of Q-disease during the first cool-down in low RRR SRF Nb cavity [Bonin, 1991]. This observation suggests that HFQS in EP cavities could be removed if the best value of the RRR is found, i.e. not too high in order to have enough number of interstitial impurities to prevent nano-hydrides formation but high enough to satisfy thermal requirement for SRF performance.

*<u>N-doped sample:</u>*

Snoek's studied the behavior of BCC metals as Nb with interstitials impurities such as O, N, C in solid solution [Snoek 1941] and he postulated a stress-induced ordering of interstitials that gives rise to pronounced internal friction peak (aka Snoek-type peaks).

As extensively explained in [Golovin 2012]: interstitial atoms (IA) within an octapore, create an elastic dipole and the lattice distortion formed has tetragonal symmetry. In the absence of external stresses IA are uniformly distributed over lattice interstices, and probability of filling an interstice within sublattices is equal. With application of external stresses, the position of IA becomes energetically favorable in octahedral interstices of one of the sublattices and IA will be transferred from one to another sublattice giving rise to the Snoek relaxation defined as the diffusion jump of IA over octahedral interstices for BCC-metals in a stress field.

Here, 2/6 N-doped samples are investigated.

As described in [Grassellino 2015; Trenhikina 2015b], two main mechanisms could be responsible for the improvement observed in N-doped samples after EP removal: 1. niobium nitride forming in layers, outer ones poorly SC and inner ones of the right composition to give NbN with $T_c$ higher than Nb; 2. residual nitrogen left as interstitial modifying the mean free path and/or the SC gap beneficially compared to 'undoped' niobium. They found that the diffraction pattern of N-doped 2/6 samples post 5 microns EP, shows only clean niobium and no nitride phases, indicating that the improved performance should be traced to low levels of interstitial (or substitutional) nitrogen rather than to good NbN phases.

In addition, by means of NED measures at cryogenic temperature they demonstrated, for the first time, precipitation of niobium hydrides in nitrogen doped cavities and samples.

Unfortunately, no investigations on N-doped cavities with different RRR values have been performed so far thus the following are pioneering results to study the effect of purity level on the N-doped treatment.

In particular, in Figure 5, N-doped spectra are compared with those coming from EP samples at low T (i.e. in the T range where Nb-H precipitation peak is expected).

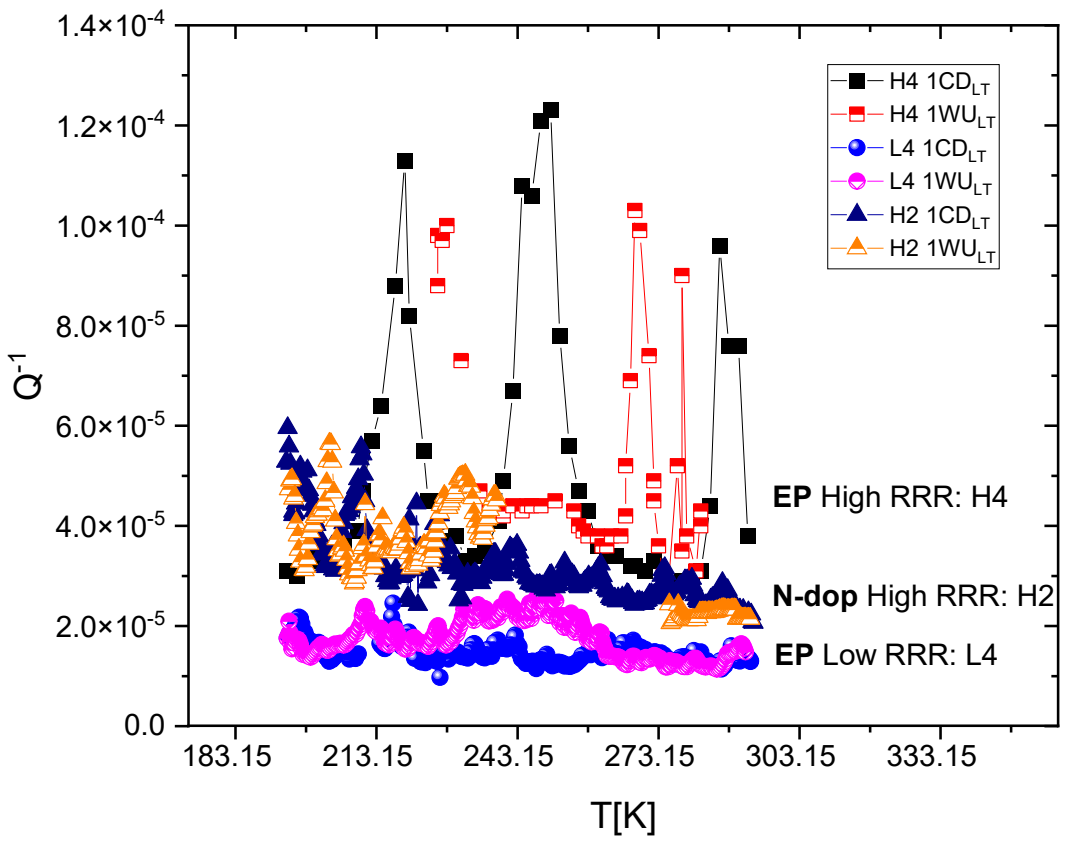

**Figure 5** Comparison N-doped vs EP samples with different RRR values – precipitation peak disappears in both EP low RRR and N-doped samples.

As for Low RRR EP sample, also for N-doped samples the small intensity of precipitation peak can be assigned to the suppression of hydrides formation due to the enhancement of trapping mechanisms by interstitial N into the BCC lattice. Indeed, as also observed by [Verdini 1980], by increasing interstitial impurities like O and N, the peak height and temperature were both reduced.

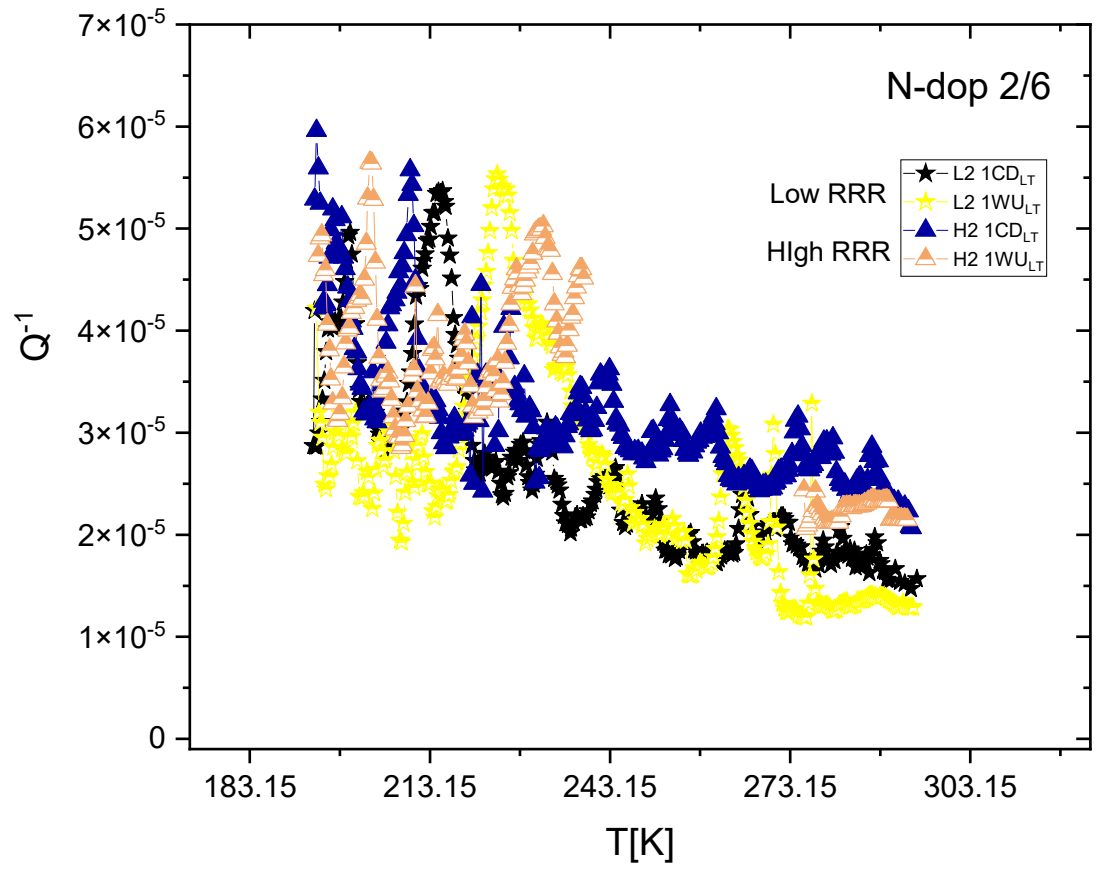

**Figure 6** Comparison N-doped samples: High RRR vs Low RRR – no differences can be found due to the poor resolution available at kHz frequencies range.

The effect of High and Low RRR on N-doped sample is also studied, see Figure 6. Both samples show similar peaks with intensity ~$4.0\times 10^{-5}$ negligible once compared with the one observed for High RRR EP samples (~1.0× $10^{-4}$). Unfortunately, due to the poor resolution (kHz) available for this study it is difficult to quantify the difference on the precipitation peak for these samples. On the other hand, we suggest that the likeness in high and low RRR N-doped is due to the presence of a comparable number of interstitial defects (i.e. comparable number of trapping sites that can prevent hydrides growth).

At higher T, as shown in Figure 7, the formation of Snoek peaks associated with N (Nb-N) and Oxygen (Nb-O) appear at ~**740K** and ~**560K** respectively.

Comparing these results with those found from literature [Alemida 2005; Florencio 2003] we observe a shift of the Snoek peak for both N and O towards higher temperature (about 100K) assigned to the higher frequency regime adopted in this internal friction study (kHz instead than Hz).

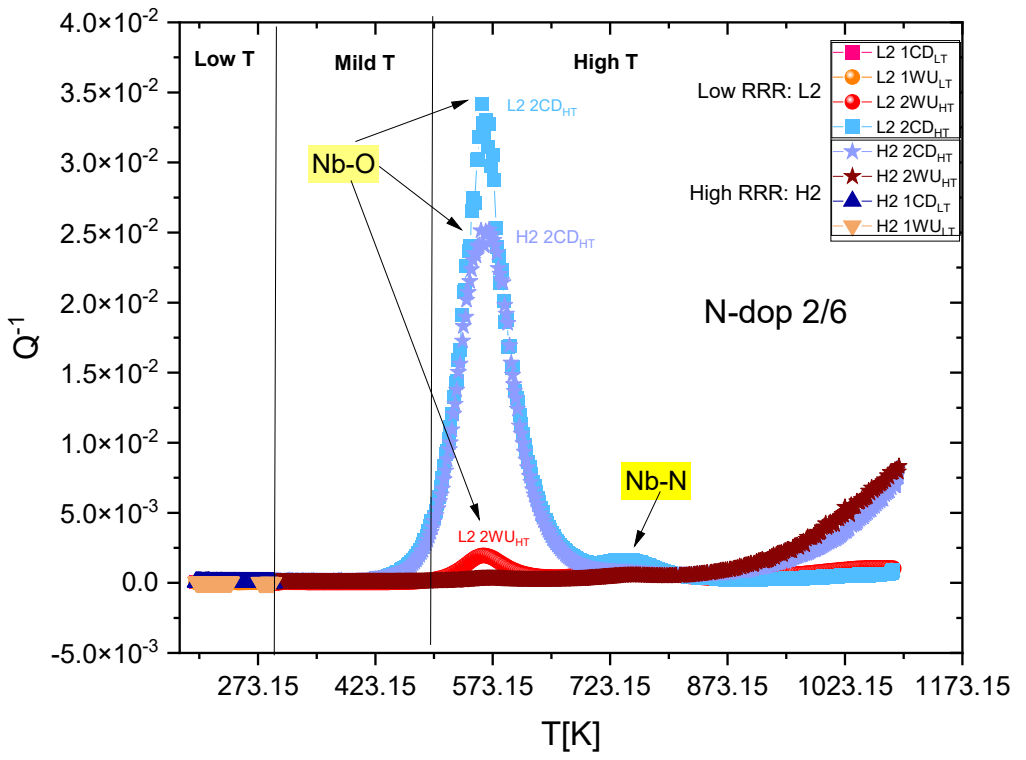

**Figure 7** Comparison N-doped samples: High RRR vs Low RRR – at high T Snoek peak due to interstitial N and O atoms appear at ~740K and ~560K.

Snoek peaks due to O appears for both N-doped samples under study (i.e., High (H2) and Low (L2) RRR) and are shown in Figure 8. It is clear that while in low RRR spectrum the Oxygen Snoek peak is already present during warm-up ($2WU_{HT}$) and its intensity increases during cool-down ($2CD_{HT}$), in high RRR sample no Snoek peak appears during warm-up (H2 $2WU_{HT}$) to then develop during cool-down from 800ºC to RT (H2 $2CD_{HT}$).

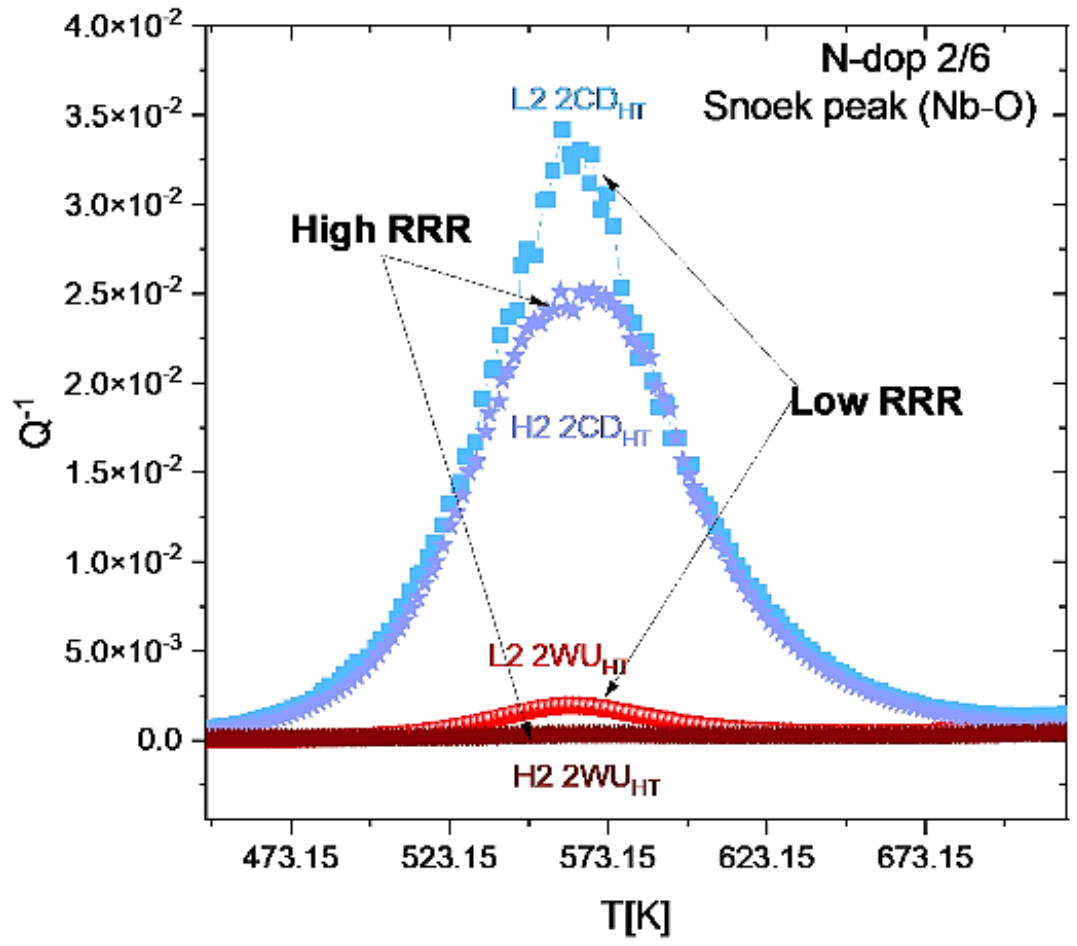

**Figure 8** N-doped samples 2/6 – Snoek peak due to Oxygen - comparison High RRR vs. Low RRR.

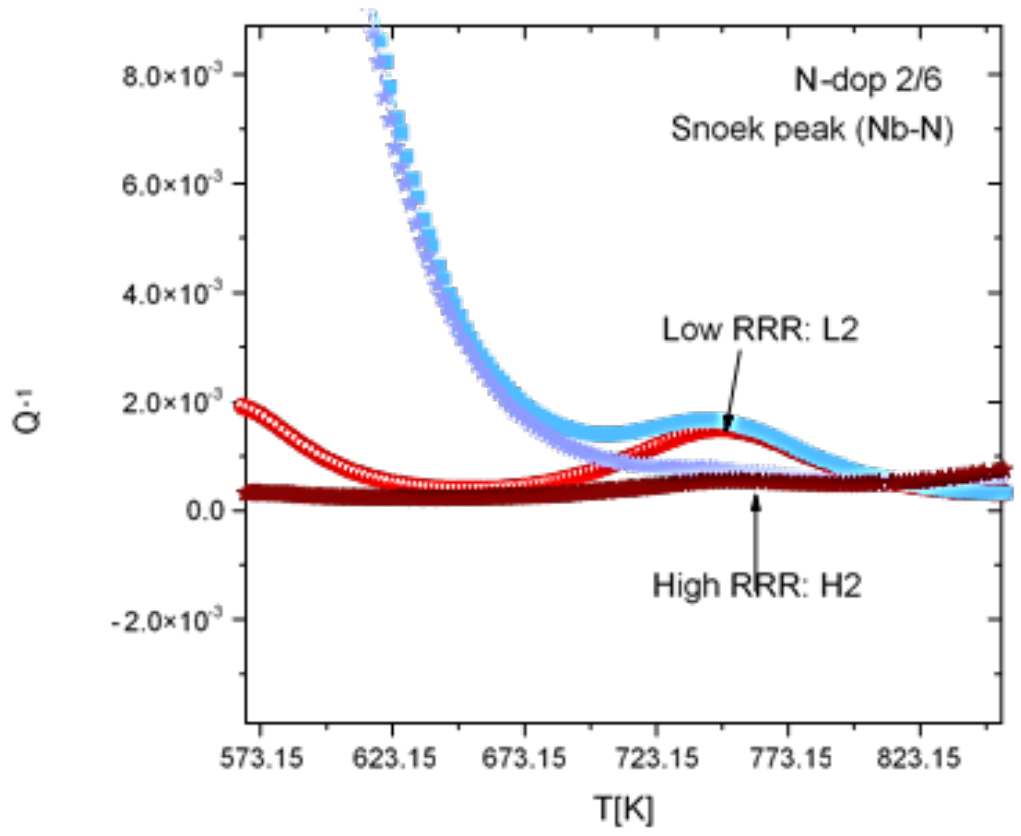

**Figure 9** N-doped samples 2/6 – Snoek peak due to Nitrogen - comparison High RRR vs. Low RRR.

This behavior can be related to the negligible amount of interstitials O typically present in High RRR sample while the appearance during CD from 800°C to RT could be assigned to the formation of an oxide layer.

For low RRR N-doped sample, comparing WU and CD at high T we also observe the enhancement in the peak intensity, and since Snoek peak is linearly proportional to the interstitial atoms in solid solution [Golovin 2012], this indicates that in this sample

Oxygen amount builds up during CD from 800°C to RT.

As shown in Figure 9, only low RRR sample shows Snoek peak due to interstitial N and its intensity doesn't change during thermal cycle meaning that there are no changes in N content during WU and CD at high T. The absence of interstitial N in high RRR samples could be assign to the formation of niobium nitrides layer but further investigations are needed to clarify this finding.

## CONCLUSIONS

In this work Internal Friction measurements have been used to explore the most important phenomena belonging to the SRF scientific community: EP and N-doping as a function of RRR.

By matching the results with those found in literature, it was proved that the suppression of hydrides is enhanced by the multiple presence of interstitial defects as vacancy and interstitial atom (O or/and N).

These results are in perfect agreement with previous anelastic relaxation theories and models, and, for the first time, Internal Friction measurements have been adopted to directly correlate surface phenomena involve in the degradation or improvement of SRF cavities.

Indeed, the highly sensitive and selective IF spectra as a function of temperature contains unique microscopic information and allows to investigate on the HFQS mitigation mechanism by N or interstitial impurities (O, vacancies) and the correlation between hydrides formation and purity level (RRR).

## ACKNOWLEDGMENTS

This work was supported by the US Department of Energy, Offices of High Energy and Nuclear Physics.

Authors would like to thanks Damon Bice for the technical assistance in the preparation of Nb samples; Dr. Bart Bollen and Dr. Nico Vanhove from IMCE (Belgium) for performing the measure of Internal Friction and above all to give us the possibility to test their apparatus IET for our purposes.

Fermilab is operated by Fermi Research Alliance, LLC under Contract No. DE-AC02- 07CH11359 with the United States Department of Energy.